# Controlling plasmonic orbital angular momentum by combining geometric and dynamic phases

Qilong Tan[1,2], Qinghua Guo[2], Hongchao Liu[2], Xuguang Huang[1,*] and Shuang Zhang[2,*]

[1]Guangzhou Key Laboratory for Special Fiber Photonic Devices and Applications, South China Normal University, Guangzhou, 510006, China

[2]School of Physics & Astronomy, University of Birmingham, Birmingham, Edgbaston, B15 2TT, United Kingdom

*Corresponding Author:
huangxg@scnu.edu.cn, s.zhang@bham.ac.uk

**Abstract:**

Tunable orbit angular momentum (OAM) of surface plasmon polaritons (SPPs) is theoretically studied with appropriately designed metasurfaces. By controlling both the orientation angle and spatial position of nano aperture array on an ultrathin gold film, the field distributions of the surface waves can be engineered to contain both spin dependent and independent OAM components. Simultaneous control over the geometric phase and optical path difference induced phase (dynamic phase) provides extra degrees of freedom for manipulating OAM of SPPs. We show that arbitrary combination of OAM numbers can be realized for the SPPs excited by incident light of different circular polarizations. The results provide powerful control over the OAM of SPPs, which will have potential applications on optical trapping, imaging, communications and quantum information processing.

## 1. Introduction

Spin-orbit interaction of light [1, 2], which describes the coupling between the spin (circular polarization) and orbital (helical phase of wavefront) degrees of freedom of photons, usually occurs when light passes through an interface or an inhomogeneous medium. In free-space, light carrying orbital angular momentum (OAM) is often referred to be an optical vortex which is expressed as a helical phase front: $e^{il\theta}$, where $l$ and $\theta$ are the topological charge and azimuth angle, respectively [3-5]. As light carrying OAM has unlimited number of eigen modes (with $l$ = 1, 2, 3. . .), it offers much more information capacity than the polarization state of light for applications in optical communications in both the classical and quantum optical regimes [6-8]. OAM can be generated with different approaches, such as diffractive optical elements [9], combination of cylindrical lenses [10], Pancheratnam-Berry phase optical elements [11], spatial light modulators [12, 13] and more recently metasurfaces [14-17]. In addition, near-field OAM are typically generated through the excitation of surface plasmon polaritons (SPPs) at the metal/dielectric interface [18, 19]. Near-field OAM has attracted considerable attention in optical trapping owing to factors such as strong field enhancement of SPPs, elimination of complicated collimation system, and ultra-compact structure [20, 21].

Existing methods to generate plasmonic OAM can be mainly divided into two categories: construction of specific phase distribution by introducing geometric phase or optical path difference induced phase (dynamic phase). The former adopts circular slits [22-24] etched on the metal film to excite plasmonic OAM by circularly polarized incident light. Spin dependent geometric phase for the excited SPPs is controlled through varying the orientation angle of each slit segment [25-27]. Arbitrary integer value of OAM number can be obtained by properly varying the orientation angle of slit segments along the azimuthal direction, however the OAM number simply reverses sign for incident light of opposite spin as the phase gradient is spin dependent. In contrast, the latter employs chiral geometries, such as spiral and whirlpool-shaped slits [28, 29] to generate additional OAM based on the optical path difference along the azimuthal angle. In contrast to the geometric phase method in which OAM number reverses sign for excitation of different spins, the difference of plasmonic OAM between right circularly polarized (RCP) and left circularly polarized (LCP) excitation is always restricted to 2 regardless of the helicity of spiral structure. This means that arbitrary OAM values and consequently higher degrees of freedom of optical manipulation using SPPs cannot be achieved by only flipping the spin of the excitation beam in previous works.

Here we propose a metasurface structure, called phase gradient spiral lens (PGSL), which consists of spatially varying nano apertures arranged on a spiral pattern carved in gold film, to generate arbitrary combinations of spin-dependent OAMs for SPPs at the metal/dielectric interface. In the PGSL design, the spin dependent and independent phases collectively contribute to the spin orbit interaction. Specific phase distribution can be constructed by introducing both geometric phase through varying the orientation angle and dynamic phase through shifting the radial position of nano apertures. Due to the unlimited control over the OAM of SPPs, it might provide new degrees of control in optical trapping [30-32] and particle rotation [33-35].

## 2. Simulation and Discussion

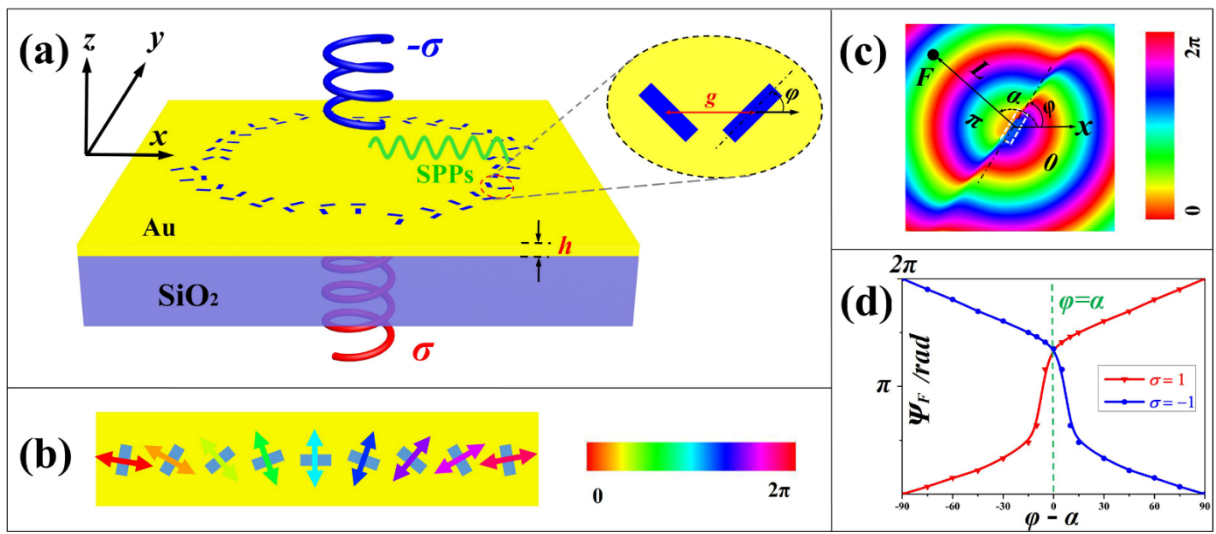

**Figure 1** Schematic of excitation of plasmonic OAMs by a metasurface. (a) A spiral array of nano apertures with spatially varying orientations on a 100 nm thick gold film excites SPPs on the gold/dielectric surface. *g* is the gap between two adjacent nano apertures in the radial direction. (b) Phase and direction of effective dipole moments for nano-apertures with different orientation angles. (c) The phase distribution of the SPPs excited by a single nano aperture. It shows an anti-phase pattern relative to the long axis of nano aperture. (d) The relation between phase of SPPs and the orientation angle at a view point $F(\alpha, L)$.

As shown in Figure 1(a), the designed metasurface consists of an array of nano apertures on a gold film with 100 nm thickness. The nano apertures with the same dimension (150 nm×400 nm) but different spatial orientations are positioned in a concentric ring or Achimedean spiral configuration. This metasurface can be easily fabricated by using Au electron-beam evaporation and focused ion-beam lithography [36].

Due to the large ratio between the transmissions of two orthogonal linear polarizations, a single nano aperture can be regarded as a local subwavelength linear polarizer and its transmitted field under normal circularly polarized illumination can be expressed as

$$E_T^\sigma = \frac{\eta}{\sqrt{2}} E_0 \exp[i\sigma(\frac{\pi}{2}+\varphi)]\begin{bmatrix}\cos\varphi\\ \sin\varphi\end{bmatrix} = \frac{\eta}{2\sqrt{2}} E_0 \left\{\begin{bmatrix}1\\ i\sigma\end{bmatrix}+e^{2i\sigma(\frac{\pi}{2}+\varphi)}\begin{bmatrix}1\\ -i\sigma\end{bmatrix}\right\} \tag{1}$$

where $\sigma = \pm 1$ is the spin of incident wave, + and - indicate the right-handed and left-handed polarization states, respectively. $E_0$ is the amplitude of EM, and $\eta$ is transmission coefficient which depends on the localized resonance of the aperture. The direction of local subwavelength linear polarizer is perpendicular to the long axis of nano aperture which is expressed as $\pi/2+\varphi$. The transmitted field of each nano aperture is linearly polarized and has a spin dependent geometric phase $\sigma\varphi$, which can be decomposed into orthogonal spin components: one preserves the same state of the incident one; the other manifests spin reverse with an additional phase of $2\sigma\varphi$, known as the Pancheratnam-Berry phase. The arrows with different colors shown in Figure 1(b) indicate the phase of the spin reversed electric field component radiation from each nano aperture with different orientation angle, whose direction reveals the local polarization of transmitted field $E_T^\sigma$.

In addition, the field resonant at each nano aperture can excite collective free electron resonances on the continuous metal/dielectric interface and form SPPs propagating along the interface with spin dependent geometric phase. The SPPs launched by each nano aperture with an orientation angle $\varphi$ carry the same additional spin dependent phase $\sigma\varphi$ as the transmitted field $E_T^\sigma$. Obviously, this spin dependent phase is limited between 0 to $\sigma\pi$ because the mirror symmetry of the nano aperture. Meanwhile, the SPP pattern excited from a nano aperture is approximately that of an in-plane point dipole and exhibits an anti-phase radiation pattern relative to the long axis of nano aperture [37], which introduces an additional fixed phase $\sigma\pi$ between the two sides of the long axis. These two phase components related to orientation angle $\varphi$ and azimuthal angle $\alpha$ both contribute to the spin dependent phase at an observation point $F$ [Figure 1(c)]. Therefore, a spin dependent phase ranged from 0 to $2\sigma\pi$ can be obtained by varying the orientation angle $\varphi$ of the nano aperture. Specifically, the phase of SPPs by a single nano aperture at the observation point $F$ can be expressed as

$$\psi_F = \sigma\varphi + \mathrm{O}(\sigma,\alpha,\varphi) - L \cdot k_{spp}, \tag{2}$$

where $α$ is the azimuthal angle of observation point $F$ relative to $x$ axis, $L$ is the distance between dipole source and observation point $F$, $k_{spp}$ is the wave vector of SPPs, and $O(σ, α, φ)$ is the phase associated with the anti-phase radiation pattern of the nano aperture. In Figure 1(d), the phase of SPPs $Ψ_F$ at a fixed observation point $F$ $(α, L)$ shows opposite trends between 0 and $2π$ by changing orientation angle $φ$. The horizontal coordinate of the intersection point depends on the azimuthal angle $α$ of the observation point $F$. As the radiation of the nano aperture is not strictly equivalent to a dipole, $O(σ, α, φ)$ does not abruptly change $π$ at the orientation angle $φ = α$, but rather shows a rapid change in the area near $φ = α$. Interestingly, it means that the orientation angle associated with anti-phase shows spin dependent shift. The rapid change area shows a different deviation from the coordinate of $φ = α$.

To achieve homogeneous amplitude of excited SPPs along the azimuthal angle, a doubling-ring nano aperture distribution is designed, as shown in Figure 1(a). The orientation angle $φ$ of outer ring has $π/2$ shift respect to the one of inner ring. In addition, the gap $g$ between two the adjacent nano apertures in the radial direction is set as odd integer multiple of $λ_{spp}/2$ which has been verified to eliminate distortion and to form a plasmonic field with a clear OAM. $λ_{spp}$ is the wavelength of SPPs with the expression as

$$λ_{spp} = λ\sqrt{\frac{\text{Re}(ε_m) + ε_d}{\text{Re}(ε_m)ε_d}} \tag{3}$$

where $λ$ is the wavelength of the excitation light in vacuum, and $ε_m$ and $ε_d$ are the permittivities of the gold and dielectric, respectively. The permittivity of gold is described by Drude model, with the plasmon frequency $ω_p = 1.37×10^{16}$ *rad/s*, and damping frequency $γ = 1.2×10^{14}$ *rad/s*. Considering the damping in propagation of SPPs, a gap $g = λ_{spp}/2$ is employed which guarantees that the field propagating towards the center along different directions is approximately uniform. Here the mutual coupling between double ring distribution nano apertures is negligible. It is worth noting that the gold film is thick enough (100 nm) to prevent the transmission of incident light at near-infrared wavelength.

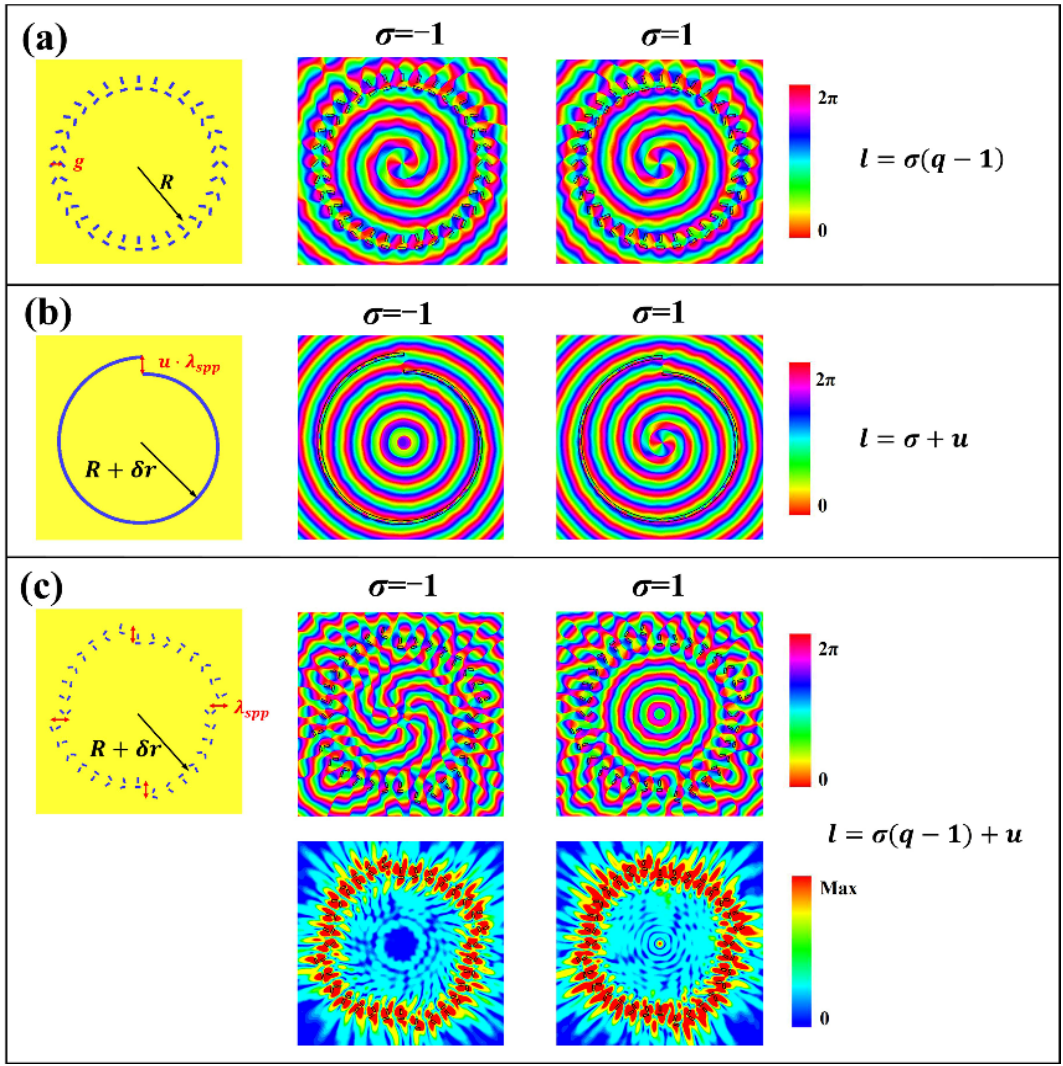

**Figure 2** Comparison between different configurations for manipulating plasmonic OAMs: (a) Nanoapertures with spatially varying orientations positioned on a circle. (b) A continuous spiral slit (c) The proposed phase gradient spiral lens. In (c), plasmonic OAM number 8 and 0 is generated corresponding to LCP ($\sigma$ = -1) and RCP ($\sigma$ = 1) incident case, respectively. Their corresponding phase and field density distributions are presented in the top and bottom rows, respectively.

To introduce spin dependent phase, Figure 2(a) shows a configuration for generating a phase distribution by varying the orientation angle of each nano aperture. The nano apertures are evenly distributed on a circle, and their orientation angles $\varphi$ varies by $q\pi$ ($q$ = 0, ±1, ±2…) across a whole turn along the circle in the clockwise direction. Each nano aperture can launch SPPs with a different initial spin dependent phase depending on its orientation angle $\varphi$. The overall phase shift of waves propagating towards the center across a whole turn equals to $2(q-1)\sigma\pi$. It can be interpreted as the combination of two parts: one is the phase shift introduced by the variation of orientation angle which is equal to $2q\sigma\pi$; the other is a phase shift associated with the anti-phase pattern which is given by $-2\sigma\pi$. As the latter is caused by a fixed change of azimuthal angle of the location of the antennas, it does not depend on $q$. Hence, the plasmonic OAM ($l$) can be expressed as $l = \sigma(q-1)$ which has been verified through the principle of electromagnetic field superposition. Although the sign of OAM number is reversed for incident light of different spins, its field intensities exhibit no difference. As shown in Figure 2(a), a case of $q = 3$ is calculated by CST Microwave Studio, and its phase distribution of vertical component of electric field ($E_z$) shows $l = \mp 2$ for LCP and RCP respectively.

The second configuration is to utilize continuous spiral or whirlpool-shaped slits for introducing a dynamic phase distribution in the azimuth direction. Besides the OAM caused by the continuous variation of the tangential orientation of the curved slit, there exists an additional OAM arising from the gradual shift along the radial direction $\delta r = u\theta\lambda_{spp}/2\pi$ ($u = 0, \pm 1, \pm 2 \ldots$) along the azimuthal angle $\theta$, where $u$ is the number of the helices. The phase gradient caused by dynamic phase can be expressed as $2\pi\delta r/\lambda_{spp}$. In contrast to the phase associated with orientation angle, this dynamic phase gradient is independent of the spin of the incident beam. As shown in Figure 2(b), as $\delta r$ gradually increases from 0 to $u\lambda_{spp}$ along the clockwise spiral slit, an additional phase gradient $2u\pi$ relative to circle ring is generated. Intuitively, the structure is equivalent to a circular ring ($q = 2$) when $u$ is equal to 0. Therefore, the plasmonic OAM number generated by spiral slit can be expressed as $l = \sigma + u$. In Figure 2(b), the case of $u = 1$ is calculated and its phase distribution of $E_z$ is obtained, showing angular momentum of 0 and 2 near the center of spiral for LCP and RCP, which leads to focus and doughnut distributions of SPPs field intensity, respectively. The difference of OAM between LCP and RCP incident case is always constrained to 2 regardless of the helix number $u$.

Both methods mentioned above are capable of generating plasmonic OAMs, but with different constraints between the OAM numbers excited by light of opposite spins. To manipulate the OAM numbers of SPP with greater freedom, the PGSL structure, as shown in Figure 1(a) and 2(c), is designed by varying the orientation angle $\varphi$ and radial shift $\delta r$ of each nano aperture simultaneously. The spatial variable ($\varphi, \delta r$) of each nano aperture is determined by both the geometric and dynamic phase, and is expressed as $\varphi(\theta) = \varphi_0 + q\theta/2$ and $\delta r(\theta) = u\theta\lambda_{spp}/2\pi$ in the polar coordinate, where $\varphi_0$ is the initial orientation angle. The spin-dependent plasmonic OAM generated by this PGSL can be expressed as $l = \sigma(q-1) + u$, which shows that arbitrary combination of OAMs for excitations of opposite spins can be readily obtained by selecting appropriate $u$ and $q$ values. As shown in Figure 2(c), a PGSL combined with controlling orientation angle ($q = -3$) and lateral shift ($u = 4$) is presented. Based on preceding analysis, its plasmonic OAM is expected to be 8 and 0 for LCP and RCP incident wave, respectively, which is confirmed by the phase distribution of the SPPs shown in Fig 2(c). In addition, the field intensity of SPPs exhibit a doughnut-like pattern for LCP and a bright focus spot right in the center for RCP excitation. The simulation results clearly demonstrate the independent and powerful control of PGSL over the plasmonic OAMs for excitations with different spins. In comparison, the two schemes represented by Fig. 2(a) and Fig. 2(b) merely correspond to two special cases of PGSL with $u = 0$ and $q = 2$, respectively.

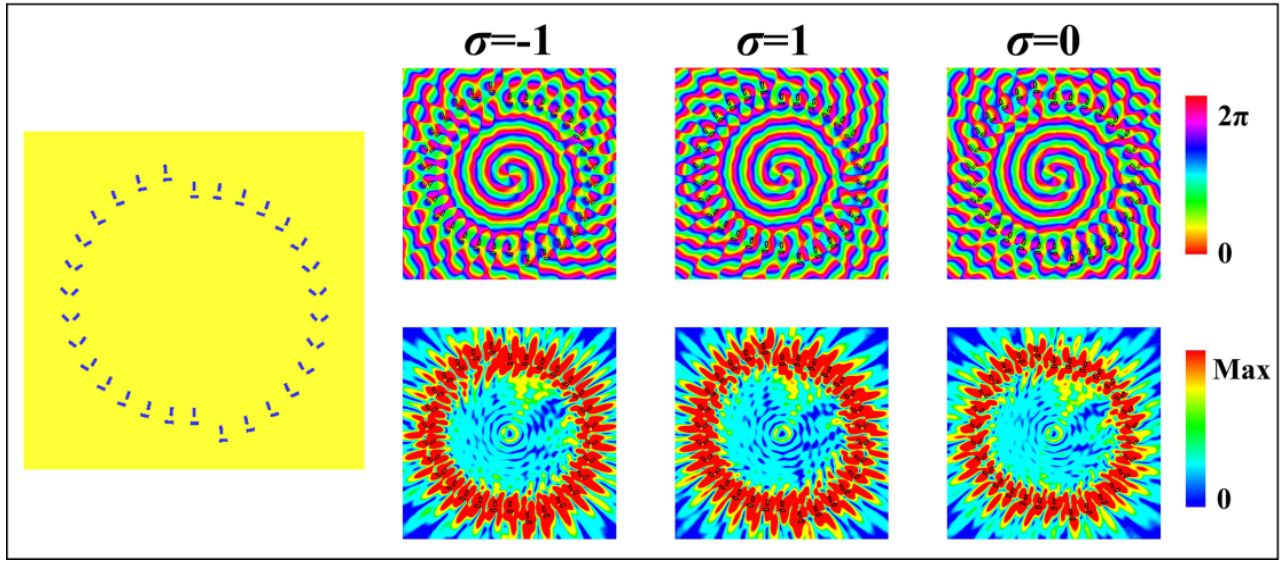

**Figure 3** Excitation of polarization independent plasmonic OAM ($q = 1$, $u = 2$). Phase distribution (top row) and field density (bottom row) of SPPs show the same OAM ($l = 2$) at the center for SPPs excited by LCP ($\sigma = -1$), RCP ($\sigma = 1$) and random linear polarization light ($\sigma = 0$).

Interestingly, for $q = 1$, there is no contribution to the plasmonic OAM from the spin dependent factor, and the total plasmonic OAM is only affected by the helix number $u$ ($l = u$), which is spin independent. Hence LCP and RCP incident light generate the same plasmonic OAM, which means that the same OAM can be excited by arbitrary polarized light. Figure 3 shows the phase distribution (top row) and field density (bottom row) of SPPs excited by LCP, RCP, and random linear polarization when the helix number is $u = 2$. An identical OAM number ($l = 2$) is obtained in all cases.

## 3. Conclusion

In conclusion, we have proposed a novel method for generating arbitrary combinations of plasmonic OAMs for incident light of opposite spins. This is achieved by combining spin-dependent geometric phase and position-dependent dynamic phase into the metasurface design through controlling both the orientation angle $\varphi$ and spatial position of each individual nano-aperture on an ultrathin gold film. Due to its powerful control over the plasmonic OAMs, potential applications on manipulating and trapping particles, florescent microscopic imaging with super resolution, and on-chip quantum information processing can be envisaged.

**Acknowledgments.** This work was supported by ERC Consolidator Grant (Topological), the EPSRC (EP/J018473/1), Leverhulme Trust (RPG-2012-674), Guangdong Natural Science Foundation (2014A030313446), the Program for Changjiang Scholars and Innovative Research Team in University (No. IRT13064), China Scholarship Council (No. 201506750035).